\documentclass[12pt]{article}

\usepackage{amsfonts}
\usepackage{epsfig}
\usepackage{a4}
\usepackage{theorem}
\usepackage{amssymb}
\usepackage{graphicx}
\input axodraw.sty

\newcommand{\be}[1]{\begin{equation}\label{#1}}
\newcommand{\ee}{\end{equation}}
\newcommand{\ba}[1]{\begin{eqnarray}\label{#1}}
\newcommand{\ea}{\end{eqnarray}}

\begin{document}
\title{Massless Three Dimensional
Quantum Electrodynamics and Thirring Model Constrained by Large
Flavor Number.}
\author{
A.R.~Fazio
\vspace{3mm} \\
\small Departamento de Fisica, Universidad Nacional de Colombia\\
\small Ciudad Universitaria, Bogot$\acute{a}$, D.C. Colombia\\
\small arfazio@unal.edu.co\vspace{3mm}\\}
\maketitle
\begin{abstract}
We explicitly prove that in three dimensional massless quantum
electrodynamics at finite temperature, zero density and large
number of flavors the number of infrared degrees of freedom is
never larger than the corresponding number of ultraviolet. Such a
result, strongly dependent on the asymptotic freedom of the
theory, is reversed in three dimensional Thirring model due to the
positive derivative of its running coupling constant.
\end{abstract}
\newpage
\section{Introduction}
A crucial question to understand a physical phenomenon is related
to the number of degrees of freedom of the physical system under
investigation. That task is not always easy because, depending on
the scale of observation of the system, degrees of freedom appear
different. Some remain coupled, some have to decouple, some are
inaccessible to our theoretical and experimental probes. It is
nice to have an universally valid constraint on quantum field
theories involving the number of the degrees of freedom of the
theory.

Some years ago, an inequality between the number of the degrees of
freedom of asymptotically free field theories in infrared and
ultraviolet regime was proposed in \cite{appelquist2} to
constraint these theories. The count of these degrees of freedom
in ultraviolet ($f_{UV}$) and in infrared ($f_{IR}$) happens
through the thermal pressure of the system, following the
definition provided later in the paper in (\ref{IR}) and
(\ref{UV}). This constraint for asymptotically free field theories
appears in form of the inequality
\begin{equation}
f_{IR}\leq f_{UV} \label{disugual},
\end{equation}
to be reversed if the theory is not asymptotically free. This
inequality has not been yet proved for a general asymptotically
free field theory, but it is consistent with the known results and
it has been used to derive new constraints for several strongly
coupled vector-like gauge theories.

In this paper we prove the validity of this inequality for
three-dimensional quantum electrodynamics (QED3) with a large
number of fermions. This theory is super-renormalizable,
UV-complete and rapidly damped at momentum scales beyond its own
mass scale. The asymptotic freedom allows for the calculation of
$f_{UV}$ like in a gas of non-interacting fermions and photons.
The existence of an infrared fixed point relies on the
calculations of the quantum corrections of $f_{IR}$, based only on
perturbation theory.

An analogous calculation is eventually performed in
three-dimensional Thirring model, weakly coupled at energy scales
much less than its own typical mass scale and in general strongly
coupled at high energy. For large number of fermions the existence
of an ultraviolet fixed point allows for the use of perturbation
theory to compute quantum corrections of $f_{UV}$ and to reverse
the inequality (\ref{disugual}).

Let us now suppose to extend the inequality (\ref{disugual}) to
finite $N$, where a strong coupling dynamics might allow for a
spontaneous symmetry breaking of some global symmetry with
corresponding dynamical mass generation. The count of degrees of
freedom involving now the Goldstone bosons, appearing in the
spectrum, might determine a critical number of fermion to generate
this mass. Intensive studies have been devoted to this critical
value of $N$ in QED3 on the continuum and on the lattice
\cite{kleinert-fiore}. It seems to be little agreement about such
a critical value, due essentially to our rudimentary understanding
of most strongly coupled quantum field theories. The constraint
(\ref{disugual}) could help in this important task.

The paper is organized as follows. In section 2 we analyze the
case of quantum electrodynamics in 2+1 dimensions. We prove that
with a large number $N$ of fermions this theory is weakly coupled
at all momentum scales. In the subsection 2.2 we compute the
thermal pressure in the large N limit and we show that our gauge
invariant result provides a negative correction to the count of
infrared degrees of freedom. Section 3 is devoted to Thirring
model in 2+1 dimensions. We compute the running of its
dimensionful coupling constant and we find an opposite behaviour
to asymptotic freedom. The bosonized equivalent version of this
model allows for an easy calculation of the quantum corrections of
the pressure and for a subsequent non-positive correction to
$f_{UV}$. Section 4 is for conclusions and perspectives.

\section{Three dimensional quantum electrodynamics}
\subsection{Large $N$ weak coupling at all momentum scale}
Let's consider $N$ massless fermions interacting with photons in
the $2+1$ dimensional Minkowsky space by the model
\begin{eqnarray}
S_{\rm QED}&=&\int d^3 x \left[\sum_{j=1}^N
\bar{\psi_j}(\imath\widehat{\partial}-
e \widehat{A})\psi_j-\frac{1}{4} F^{\mu\nu}F_{\mu\nu}\right]\\
(\widehat{\imath\partial}- e
\widehat{A})&=&\gamma^\mu(\imath\partial_\mu- e
A_\mu)\nonumber\\
F_{\mu\nu}&=&\partial_\mu A_\nu-\partial_\nu A_\mu \nonumber
\end{eqnarray}
The squared of the coupling constant $e$ has dimension of a mass
and the interaction is super-renormalizable. We will be working in
the large $N$ limit by keeping fixed the mass $\frac{e^2 N}{8}$.
It is remarkable that $\psi_j$ is a set of $N$ 4-component fermion
fields. This sort of redoubling of the fermions number avoids any
Chern-Simon terms in the Lagrangian \cite{templeton-delbourgo}.

As a first step let's prove that for large $N$ the theory remains
weakly coupled at all momentum scales. In a general covariant
gauge the leading quantum correction to the gauge boson propagator
of momentum $p$ is
\begin{equation}
Ne^2\left[\frac{ g_{\mu\rho}}{p^2}-(1-\xi)\frac{p_\mu
p_\rho}{p^4}\right]\int \frac{d^3 q}{(2\pi)^3}
Tr\left(\gamma^\rho\frac{1}{\widehat
q}\gamma^\sigma\frac{1}{\widehat {q-p}}\right)\left[\frac{
g_{\sigma\nu}}{p^2}-(1-\xi)\frac{p_\sigma p_\nu}{p^4}\right]
\end{equation}
After cutting the external legs and making some $\gamma$ gymnastic
one easily obtains
\begin{equation}
4N e^2\int \frac{d^3 q}{(2\pi)^3}\frac{1}{q^2 (q - p)^2}(2q^\rho
q^\sigma - g^{\rho\sigma}q^2-q^\rho p^\sigma -p^\rho q^\sigma
+g^{\rho\sigma}q \cdot p). \label{prop1}
\end{equation}
The use of Feynman's parameters implies \cite{collins}
\begin{eqnarray}
\frac{1}{q^2 (q-p)^2}&=&\int_0^1 dx \frac{1}{(-q^2+2x q\cdot p-x
p^2)^2}\\
\int \frac{d^3 q}{(2\pi)^3}\frac{q_\rho q_\sigma}{q^2
(q-p)^2}&=&\frac{1}{64\sqrt{p^2}}(3p_\rho p_\sigma - g_{\rho\sigma}p^2)\\
\int \frac{d^3 q}{(2\pi)^3}\frac{q_\rho}{q^2
(q-p)^2}&=&\frac{p_\rho}{16\sqrt{p^2}}.
\end{eqnarray}
Collecting all the previous results the photon propagator till to
one loop is given, in a general $\xi$-covariant gauge, by
\begin{equation}
-\frac{\imath}{p^2}\left(g_{\mu\nu}-\frac{p_\mu
p_\nu}{p^2}\right)\left(1-\frac{Ne^2}{8\sqrt{-p^2}}\right)-\imath\xi\frac{p_\mu
p_\nu}{p^4}\label{propagatore},
\end{equation}
which is an exact result to leading order in the $1/N$ expansion,
showing that only the transverse gauge independent part is
renormalized by interactions.

It is worth to remark that due to the dimensionality of the
space-time no infrared divergences are encountered. In spite of
superficial ultraviolet divergences, they too are absent. In fact,
in (\ref{prop1}) the only possible divergent term is
\begin{equation}
\int \frac{d^3 q}{(2\pi)^3}\frac{1}{q^2}.
\end{equation}
That is zero, as it is easy to see by using the Feynman parameter
$x$ in
\begin{equation}
\int \frac{d^3 q}{(2\pi)^3}\frac{1}{q^2}\frac{(p+q)^2}{(p+q)^2}
\end{equation}
which reduces to
\begin{equation}
\imath\frac{\pi^{3/2}}{(2\pi)^3}\Gamma\left(\frac{1}{2}\right)\sqrt{p^2}\int_0^1
(x^2-x)^{-\frac{1}{2}}(4x^2-3x)=0\label{prop2}.
\end{equation}
This correction amounts to extract from the gauge boson propagator
(\ref{propagatore}) a dimensionless coupling running constant
\cite{heinz}
\begin{equation}
\bar{\alpha}(p^2) = \frac{e^2 \sqrt{-p^2}}{-p^2+\frac{N
e^2}{8}\sqrt{-p^2}}
\end{equation}
which in the Euclidean three-dimensional space-time appears as
\begin{equation}
\bar{\alpha}(p) = \frac{e^2 p}{p^2+\frac{N e^2}{8}p}.
\label{coupling}
\end{equation}
This expression exhibits asymptotic freedom at large momentum and
conformal symmetry with an infrared fixed point of strength
$\frac{8}{N}$ as $p$ tends to zero. For large $N$ the coupling is
always weak and we assume no dynamical fermion mass generation.
\subsection{Appelquist, Cohen, Schmaltz's conjectured inequality}
In \cite{appelquist2} and \cite{appelquist1} a constraint on this
theory in the large flavor number was conjectured in form of an
inequality stating that the number of infrared degrees of freedom
$f_{IR}$, defined using the thermal pressure $P(T)$ at the
absolute temperature $T=\frac{1}{\beta}$ of the above interacting
fermions and photon gas, is never larger than the number of the
corresponding ultraviolet degrees of freedom. To be more specific
\begin{equation}
f_{IR}=\lim_{T \rightarrow 0}\frac{P(T)f(d)}{T^d} \label{IR}
\end{equation}
and
\begin{equation}
f_{UV}=\lim_{T \rightarrow +\infty}\frac{P(T)f(d)}{T^d},
\label{UV}
\end{equation}
where $P(T)$ is related to the grand canonical partition function
$Z$ as
\begin{equation}
P(T)=\frac{T \log Z}{V} \label{press.}
\end{equation}
and $f(d)$ is a function of the number of Euclidean space-time
dimensions $d$, defined such that the contribution from a free
bosonic field is equal to 1, amounting therefore to:
\begin{equation}
f(d)= 2^{d-1}\pi^{\frac{d-1}{2}}\Gamma \left(\frac{d+1}{2}\right)
\times\frac{1}{\Gamma(d)\zeta(d)}.
\end{equation}
It is a straightforward exercise to compute the above numbers of
degrees of freedom of a gas of non-interacting fermions and
non-interacting photons \cite{Kapusta}. In $2+1$ dimensions one
obtains that the $4N$ fermion components are Boltzman-weighted by
a $3/4$ factor.

Due to the above proved asymptotic freedom the result for $f_{UV}$
is
\begin{equation}
f_{UV}=\frac{3}{4}(4N)+1 \label{free}
\end{equation}
where we have taken into account that in three space-time
dimensions only one degree of freedom per gauge boson field is
propagating.

The leading contribution to $f_{IR}$ is given by (\ref{free}), but
due to the weak coupling (\ref{coupling}) it can be corrected
using perturbation theory. The lowest order correction to $f_{IR}$
due to interactions comes from the usual two loop exchanging term
contribution to the pressure $P(T)$ (\ref{press.}), which in the
formalism of the imaginary time \cite{kapusta2} is
\begin{equation}
-\frac{1}{2}\,N\,T\,\bar{\alpha}(T)\,\int\frac{d^2
p}{(2\pi)^2}\int\frac{d^2 q}{(2\pi)^2}\int\frac{d^2 k}{(2\pi)^2}
(2\pi)^2 \delta^2(\vec{p}-\vec{q}-\vec{k})
\sum_{n_p,n_q,n_k}\beta\delta_{n_p,n_q+n_k}T^3\frac{Tr(\gamma^\mu
\widehat{p}\gamma_\mu\widehat{q})}{p^2 q^2 k^2}. \label{2loop}
\end{equation}
In the analytic expression of this contribution we took into
account the $\frac{1}{N}$ resummation using $T\,\bar{\alpha}(T)$
as perturbative parameter and the free gauge boson propagator in
(\ref{2loop}).

By simply dimensional analysis the contribution (\ref{2loop}) is
proportional to $N\,\bar{\alpha}(T)\,T^3$ and, since
$$\lim_{T\rightarrow 0}\bar{\alpha}(T)=\frac{8}{N},$$
it is expected to be $O(1)$ in this large N perturbative
expansion. It is easy to check that higher corrections are
suppressed like $\frac{1}{N}$ powers.

The calculation of the above diagram is performed in Feynman
gauge, eventually we give an argument in favour of the gauge
invariance of our result. The simplicity of the calculations in
Feynman gauge, already exploited in four dimensions
\cite{kapusta2}, will be recovered in our case, where the task
will be simplified due to the absence of nasty subdivergences.

Firstly let us make the sum on the Matsubara frequencies
\begin{equation}
\sum_{n_p,n_q,n_k}\beta\delta_{n_p,n_q+n_k}T^3\frac{Tr(\gamma^\mu
\widehat{p}\gamma_\mu\widehat{q})}{p^2 q^2 k^2}. \label{mats}
\end{equation}
Due to the periodic and antiperiodic boundary conditions for,
respectively, gauge bosons and fermions one can perform an
analytic continuation by writing
$$\beta\delta_{n_p,n_q+n_k}=\int_0^\beta d\theta\,\, exp[\theta(p^0-q^0-k^0)]
$$
and multiplying (\ref{mats}) by the quantity
$$-exp[\beta(k^0+q^0)]$$
without making any change. The above Matsubara sum becomes
\begin{equation}
4\,T\Sigma_{n_k}\,\frac{1}{k^2}\,\,T\Sigma_{n_p}\,\frac{1}{p^2}\,\,T\Sigma_
{n_q}\,\frac{1}{q^2}\,\,\frac{p\cdot q}{p^0-q^0-k^0}\{exp(\beta
p^0)-exp[\beta(k^0+q^0)]\}. \label{mats1}
\end{equation}
The contour integral result for the $n_p$ summation is
\begin{equation}
T\Sigma_{n_p}\frac{I(p^0,q^0,k^0)}{p^2}=\frac{I(E_p,q^0,k^0)}{2E_p}N_F(p)
+\frac{I(-E_p,q^0,k^0)}{2E_p}(N_F(p)-1)
\end{equation}
where $I(p^0,q^0,k^0)$ is an arbitrary analytical function and
$$E_p=|\vec{p}|\,\,\,\,\,\,\,\,\,\,\,\,\,\,\,\,\,\, N_F(p)=\frac{1}{1+exp(\beta E_p)};$$

the $n_k$ summation amounts to
\begin{equation}
T\Sigma_{n_k}\frac{I(p^0,q^0,k^0)}{k^2}=-\frac{I(p^0,q^0,E_k)}
{2E_k}N_B(k) -\frac{I(p^0,q^0,-E_k)}{2E_k}(N_B(k)+1)
\end{equation}
where
$$E_k=|\vec{k}|\,\,\,\,\,\,\,\,\,\,\,\,\,\,\,\,\,\,
N_B(k)=\frac{1}{exp(\beta E_k)-1}.$$

Taking in particular
\begin{equation}
I(p^0,q^0,k^0)= \frac{p\cdot q}{p^0-q^0-k^0}\{exp(\beta
p^0)-exp[\beta(k^0+q^0)]\}
\end{equation}
our Matsubara frequencies sum (\ref{mats1}) can be written as
\begin{eqnarray}
&&-\frac{1}{2 E_k E_p
E_q}\{N_F(p)N_F(q)[I(E_p,E_q,-E_k)(N_B(k)+1)+I(E_p,E_q,E_k)N_B(k)]+\nonumber\\
&&N_F(q)(N_F(p)-1)[I(-E_p,E_q,-E_k)(N_B(k)+1)+I(-E_p,E_q,E_k)N_B(k)]+\nonumber\\
&&(N_F(q)-1)N_F(p)[I(E_p,-E_q,-E_k)(N_B(k)+1)+I(E_p,-E_q,E_k)N_B(k)]+
\\
&&(N_F(p)-1)(N_F(q)-1)[I(-E_p,-E_q,-E_k)(N_B(k)+1)+I(-E_p,-E_q,E_k)
N_B(k)]\}.\nonumber
\end{eqnarray}
After some algebra this result can be simplified to
\begin{eqnarray}
&&-\frac{1}{E_k E_p E_q}\Big\{\frac{E_p
E_q-\vec{p}\cdot\vec{q}}{E_p-E_q+E_k}(N_F(q)+N_B(k)N_F(q)-N_F(p)N_F(q)
-N_B(k)N_F(p))\nonumber\\
&&+\frac{E_p
E_q-\vec{p}\cdot\vec{q}}{E_p-E_q-E_k}(N_F(p)N_F(q)+N_B(k)N_F(q)
-N_F(p)N_B(k)-N_F(p))\nonumber\\
&&+\frac{E_p
E_q+\vec{p}\cdot\vec{q}}{E_p+E_q-E_k}(N_B(k)-N_B(k)N_F(q)-N_F(p)N_F(q)
-N_B(k)N_F(p))\\
&&+\frac{E_p
E_q+\vec{p}\cdot\vec{q}}{E_p+E_q+E_k}(1+N_B(k)-N_F(q)-N_F(p)
-N_B(k)N_F(q)+N_F(p)N_F(q)-N_B(k)N_F(p))\}.\nonumber
\end{eqnarray}
Since the set of values of the physical thermal pressure has the
algebraic structure of a torsor, the temperature independent
vacuum term can be neglected. The linear terms in the occupation
number provide a vanishing contribution to the pressure by
symmetric integration and using the result (\ref{prop2}), which
implies
\begin{equation}
\int\frac{d^2 p}{(2\pi)^2}\frac{1}{2E_q}=0.
\end{equation}

The final expression for our exchange term (\ref{2loop}) is
\begin{equation}
-3\,N\, T\bar{\alpha}(T) \int\frac{d^2 p}{(2\pi)^2}\int\frac{d^2
q}{(2\pi)^2}\int\frac{d^2 k}{(2\pi)^2} (2\pi)^2
\delta^2(\vec{p}-\vec{q}-\vec{k})\frac{N_F(p)}{E_p}\frac{N_F(q)}{E_q}
\end{equation}
and since
\begin{equation}
\int\frac{d^2
q}{(2\pi)^2}\frac{N_F(p)}{E_p}=\frac{1}{2\pi}\int_0^{+\infty} dx
\frac{1}{e^{\beta x}+1}=\frac{\log 2}{2\pi \beta}
\end{equation}
the one-loop corrected $f_{IR}$ (\ref{IR}) is
\begin{equation}
f_{IR}=3N+1-\frac{12\,\, log^2 2}{\pi\zeta(3)}
\end{equation}
confirming the conjectured inequality
\begin{equation}
f_{IR}\leq f_{UV} \label{conject}
\end{equation}
in Feynman gauge.

Since this result is based on the calculation of a physical
observable as the thermal pressure, one can be convinced about its
gauge independence, also if it is not so easy to provide an
explicit proof about that. Let's say that in non-covariant gauges
as the Coulomb and temporal gauge the pressure was computed for
QCD in $3+1$ dimensions in \cite{Shuryak} obtaining the same
result as in Feynman gauge, therefore an analogous invariance is
expected for QED3.

In a general $\xi$ covariant gauge one should add to the simple
expression (\ref{2loop}) the term
\begin{eqnarray}
&&\frac{(1-\xi)}{2}N T \bar{\alpha}(T)\int\frac{d^2
p}{(2\pi)^2}\int\frac{d^2 q}{(2\pi)^2}\int\frac{d^2 k}{(2\pi)^2}
(2\pi)^2 \delta^2(\vec{p}-\vec{q}-\vec{k})\times\nonumber\\
&&\sum_{n_p,n_q,n_k}\beta\delta_{n_p,n_q+n_k}T^3\frac{1}{p^2 q^2
k^4} Tr(\widehat{k}\, \widehat{p}\,\widehat{k}\,\widehat{q}).
 \label{2Lloop}
\end{eqnarray}
The calculation of (\ref{2Lloop}) proceeds in the same way as in
Feynman gauge, although the contour integral method involves now a
double pole residue, providing the Matsubara frequency sum on
$n_k$
\begin{eqnarray}
\frac{1}{\beta}\Sigma_{n_k}\frac{1}{(k^2)^2}I(p^0,q^0,k^0)&=&
-\frac{N_B(k)}{4E_k^2}\frac{\partial I}{\partial
k^0}(p^0,q^0,E_k)+\frac{N_B(k)+1}{4E_k^2}\frac{\partial
I}{\partial
k^0}(p^0,q^0,-E_k)\nonumber\\&&+\frac{\beta(N_B(k)+N_B(k)^2)}
{4E_k^2}I(p^0,q^0,E_k)+
\frac{\beta(N_B(k)+N_B(k)^2)}{4E_k^2}I(p^0,q^0,-E_k)\nonumber\\
&&+\frac{N_B(k)I(p^0,q^0,E_k)}{4E_k^3}+\frac{(N_B(k)+1)I(p^0,q^0,-E_k)}{4E_k^3}
\end{eqnarray}

However the term (\ref{2Lloop}) doesn't contribute to the count of
infrared degrees of freedom of the three dimensional quantum
electrodynamics, as it can be seen by the following simple
argument\cite{Bjorken}. Let' write (\ref{2Lloop}) as
\begin{eqnarray}
&&\frac{(1-\xi)}{2}N T \bar{\alpha}(T)\int\frac{d^2
p}{(2\pi)^2}\int\frac{d^2 q}{(2\pi)^2}\int\frac{d^2 k}{(2\pi)^2}
(2\pi)^2 \delta^2(\vec{p}-\vec{q}-\vec{k})\times\nonumber\\
&&T^3\sum_{n_p,n_q,n_k}\beta\delta_{n_p,n_q+n_k}
\frac{1}{k^4}Tr\left(\widehat{k}\,
\frac{1}{\widehat{p}}\,\widehat{k}\,\frac{1}{\widehat{q}}\right).
\label{2Lloop1}
\end{eqnarray}
Integration on $\vec{p}$ and summation on $n_p$ reduce this
integral to
\begin{equation}
\frac{(1-\xi)}{2}N T \bar{\alpha}(T)\int\frac{d^2
q}{(2\pi)^2}\int\frac{d^2 k}{(2\pi)^2}
T^2\sum_{n_q,n_k}\frac{1}{k^4}Tr\left(\widehat{k}
\frac{1}{\widehat{k+q}}\widehat{k}\frac{1}{\widehat{q}}\right),
\label{2Lloop2}
\end{equation}
but
\begin{equation}
\frac{1}{\widehat{k+q}}\widehat{k}\frac{1}{\widehat{q}}=
\frac{1}{\widehat{q}}-\frac{1}{\widehat{k+q}}
\end{equation}
and due to the convergence of (\ref{2Lloop}) which guarantees the
shifting of the origin in $q$-momentum space, the integral
(\ref{2Lloop}) vanishes.
\section{Thirring Model}
The $2+1$ dimensional action for Thirring model with a large
number $N$ of fermions is:
\begin{equation}
S=\int d^3\,x\,\left[\sum_{j=1}^{N}\bar{\psi}_j\imath
\widehat{\partial}\psi_j-\frac{g^2}{2}\sum_{j=1}^{N}
(\bar{\psi}_j\gamma_\mu\psi_j)^2\right] \label{Thirr}
\end{equation}
The coupling constant $g^2$ has dimension of a inverse of a mass,
and the theory is nonrenormalizable. However it has been proved
\cite{hands} that in $2+1$ dimensions no ultraviolet divergences
appear in a scheme preserving the current conservation, if an
expansion in the dimensionless parameter $1/N$ is performed.
Supported by this result we claim that in the three dimensional
Euclidean space-time it is possible to extract from the 4-point
Green function of this model a dimensionless running coupling
constant
\begin{equation}
\tilde{g^2}(p)=\frac{g^2 p}{1+\frac{N}{8}g^2p}.
\label{couplingThirr}
\end{equation}
It shows that the model becomes non-interacting in the infrared
($g^2 p\ll 1$) and it has a weak ultraviolet fixed point at
$\frac{8}{N}$, in strength exactly equal to the infrared one of
three dimensional quantum electrodynamics. This result is in
perfect agreement with that one obtained in \cite{hands}, working
in the bosonized version of the model, whose Lagrangian in the
three dimensional Euclidean space is
\begin{equation}
\textsf{L}=\sum_{j=1}^{N}\bar{\psi}_j\widehat{\partial}\psi_j+\imath
g\sum_{j=1}^{N}\bar{\psi}_j\widehat{A}\psi_j+\frac{1}{2}{A_\mu}{A_\mu}
\label{bosonized}
\end{equation}
the field $A_\mu$ is an auxiliary vector; it may be integrated
over to recover the original model (\ref{Thirr}). To leading order
in $1/N$ the auxiliary propagator receives a contribution from
vacuum polarization, so we write
\begin{equation}
D_{\mu\nu}(p)= \left(\delta_{\mu\nu}-\frac{p_\mu
p_\nu}{p^2}\right)\frac{1}{1+\frac{N}{8}g^2p}+\frac{p_\mu
p_\nu}{p^2}.
\end{equation}

Extending the Lagrangian (\ref{bosonized}) at finite temperature,
it is possible to extract the physical degrees of freedom
following the definitions (\ref{IR}) and (\ref{UV}). At the
leading order one finds:
\begin{equation}
f_{IR}=f_{UV}=3N,
\end{equation}
then $f_{UV}$ can be corrected in perturbation theory. The
calculation of $f_{UV}$ next-to-leading correction proceeds in the
same way as in QED. Up to a positive proportionality factor that
correction is
\begin{equation}
\frac{N\tilde{g^2}(T)}{2 T^4}\int\frac{d^2
p}{(2\pi)^2}\int\frac{d^2 q}{(2\pi)^2}\int\frac{d^2 k}{(2\pi)^2}
(2\pi)^2 \delta^2(\vec{p}-\vec{q}-\vec{k})
\sum_{n_p,n_q,n_k}\beta\delta_{n_p,n_q+n_k}T^3\frac{Tr(\gamma^\mu
\widehat{p}\gamma_\mu\widehat{q})}{p^2 q^2}.\label{2loopthirring}
\end{equation}
The summation on the Matsubara frequencies is now highly
simplified, the result is
\begin{eqnarray}
&&T^3\sum_{n_p,n_q,n_k}\beta\delta_{n_p,n_q+n_k}\frac{Tr(\gamma^\mu
\widehat{p}\gamma_\mu\widehat{q})}{p^2
q^2}=\nonumber\\&&\frac{1}{\beta E_p
E_q}\left[(N_F(p)-N_F(q))\frac{E_p E_q-\vec{p}\cdot
\vec{q}}{E_p-E_q}+(N_F(p)+N_F(q))\frac{E_p E_q+\vec{p}\cdot
\vec{q}}{E_p+E_q}\right]. \label{2loopthirring1}
\end{eqnarray}
Substituting (\ref{2loopthirring1}) into (\ref{2loopthirring}) we
obtain
\begin{equation}
\frac{2\pi N\tilde{g^2}(T)}{\beta T^4}\int_0^{+\infty}d E_p d
E_q\,\, E_p E_q\,\,\frac{E_p N_F(q)- E_q N_F(p)}{E_p^2-E_q^2}.
\label{last}
\end{equation}
But (\ref{last}) is convergent and
\begin{equation}
\forall E_p,E_q\,\,\,\,\,\frac{E_p N_F(q)- E_q
N_F(p)}{E_p^2-E_q^2}<0,
\end{equation}
moreover
\begin{equation}
\lim_{T\rightarrow +\infty}N\tilde{g^2}(T)= 8
\end{equation}
we therefore conclude that the $f_{UV}$ next-to-leading correction
is non-positive and
\begin{equation}
f_{UV}\leq f_{IR}.
\end{equation}
\section{Conclusions and perspectives}
The existence of a conjectured inequality between infrared and
ultraviolet degrees of freedom has been explicitly verified in
QED3 and in 3D-Thirring model with large $N$ number of fermions.
The calculations have been done in the formalism of imaginary time
with large $N$ limit resummation and gauge invariant results have
been obtained.

In a subsequent paper we plan to verify this inequality when the
global symmetry $U(N)\bigotimes U(N)$ of QED3 is spontaneously
broken and a non-linear realization of that symmetry is working at
low energy through $2N^2$ Goldstone bosons and massive fermions.
In that case the expected critical number of fermions for
dynamical mass generation is $\frac{3}{2}$. As it was pointed out
in \cite{appelquist1} a natural question about the application of
$f_{IR}\leq f_{UV}$ in the context of spontaneous symmetry
breaking of QED3 has to do with the Mermin-Wagner-Coleman theorem
\cite{MWC}, stating that spontaneous symmetry breaking of a
continuous symmetry cannot happen in $2+1$ dimensions, as for
example it has been verified in the renormalizable theory $\lambda
\varphi^6$ \cite{anaos}. It is in fact expected \cite{RWP} that at
small temperature Goldstone bosons acquire masses, but it would
not affect the count of infrared degrees of freedom (\ref{IR}). An
explicit computation in this sense is in progress.

Following the recent proposal of \cite{sannino} we are
investigating about the implications of this inequality, in
quantum chromo-dynamics too. Since the analytic results about the
condensate quark-antiquark in orbifold QCD \cite{ASV} it would be
interesting to see if a critical number of fermions, compatible
with our inequality, could be extracted from the analytic value of
this condensate.

\section*{Acknowledgments}
The author would like to thank, for support and invitation, the
organizers of the XVIth Indian Summer School in Prague, where he
had the possibility to learn about this new constraint in a series
of lectures given by Prof. T. Appelquist. The School of Physics of
the University of the Witwatersrand (Johannesburg), where this
investigation started, is acknowledged too.

\end{document}